%% file: main.tex
\documentclass[10pt,conference]{IEEEtran}
\IEEEoverridecommandlockouts
\usepackage{cite}
\usepackage{amsmath,amssymb,amsfonts}
\usepackage{graphicx}
\usepackage{textcomp}
\usepackage{xcolor}
\usepackage[normalem]{ulem}

\def\BibTeX{{\rm B\kern-.05em{\sc i\kern-.025em b}\kern-.08em
    T\kern-.1667em\lower.7ex\hbox{E}\kern-.125emX}}

\input{local}

\begin{document}

\title{\tool: Improving Secure Code Review with Mixture of Prompts}

\author{
\IEEEauthorblockN{Yun Peng\IEEEauthorrefmark{2}, Kisub Kim\IEEEauthorrefmark{3}\IEEEauthorrefmark{1}, Linghan Meng\IEEEauthorrefmark{4}, Kui Liu\IEEEauthorrefmark{4}}
\IEEEauthorblockA{\IEEEauthorrefmark{2}The Chinese University of Hong Kong}
\IEEEauthorblockA{\IEEEauthorrefmark{3}DGIST, Daegu, Korea}
\IEEEauthorblockA{\IEEEauthorrefmark{4}Huawei Technologies, China}
\IEEEauthorblockA{ypeng@cse.cuhk.edu.hk falconlk00@gmail.com \{menglinghan2, kui.liu\}@huawei.com}
}

\maketitle

\input{sections/abs}
\input{sections/intro}
\input{sections/mot}
\input{sections/meth}
\input{sections/setup}
\input{sections/eval}
\input{sections/discussion}
\input{sections/literature}
\input{sections/conclusion}

\bibliographystyle{plain}
\bibliography{ref}

\end{document}

%% file: local.tex
\usepackage{framed}
\usepackage[T1]{fontenc}
\usepackage{listings}
\usepackage{xcolor}
\usepackage{color}
\usepackage{multirow}
\usepackage{algorithm}

\usepackage{amsmath}
\usepackage{amssymb}
\usepackage{threeparttable}
\usepackage{url}
\usepackage[most]{tcolorbox}
\usepackage{hyperref}
\usepackage[hyphenbreaks]{breakurl}
\usepackage{algpseudocode}
\usepackage{colortbl}
\usepackage{booktabs}
\usepackage{balance}
\hypersetup{
    colorlinks = true,
    linkcolor=blue,
    filecolor=blue,      
    urlcolor=blue,
    citecolor=cyan,
}
\usepackage{xspace}
\usepackage{subfigure}
\usepackage{graphicx}
\usepackage{fvextra}
\usepackage{pifont}

\usepackage[shortlabels]{enumitem}
\setlist[enumerate]{nosep}

\definecolor{codegreen}{rgb}{0,0.6,0}
\definecolor{codegray}{rgb}{0.5,0.5,0.5}
\definecolor{codepurple}{rgb}{0.58,0,0.82}
\definecolor{backcolour}{rgb}{0.95,0.95,0.92}
\definecolor{lightgreen}{rgb}{0,0.4,0}
\definecolor{lightred}{rgb}{0.4,0,0}

\definecolor{mygray}{gray}{.9}

\lstdefinestyle{mystyle}{
    commentstyle=\color{brown},
    keywordstyle=\color{magenta},
    numberstyle=\tiny\color{codegray},
    stringstyle=\color{codepurple},
    basicstyle=\ttfamily\footnotesize,
    breakatwhitespace=false,         
    breaklines=true,                 
    captionpos=b,                    
    keepspaces=true,                 
    numbers=left,                    
    numbersep=5pt,                  
    showspaces=false,                
    showstringspaces=false,
    showtabs=false,                  
    tabsize=2,
    escapeinside={<@}{@>}
}

\newcommand{\tool}{\text{iCodeReviewer}\xspace}

\newcommand\etal{{\it{et al.\ }}}

\newcommand{\tabincell}[2]{\begin{tabular}{@{}#1@{}}#2\end{tabular}}
\newcommand{\eat}[1]{\if 0 #1 \fi}

\newfont{\mycrnotice}{ptmr8t at 7pt}
\newfont{\myconfname}{ptmri8t at 7pt}

\newcommand{\answer}[2]{
\begin{tcolorbox}[breakable,width=\linewidth,boxrule=0pt,top=1pt, bottom=1pt, left=1pt,right=1pt, colback=gray!20,colframe=gray!20]
\textbf{Answer to RQ#1:} #2
\end{tcolorbox}
}

%% file: sections/abs.tex
\begin{abstract}
    Code review is an essential process to ensure the quality of software that identifies potential software issues at an early stage of software development. Among all software issues, security issues are the most important to identify, as they can easily lead to severe software crashes and service disruptions. Recent research efforts have been devoted to automated approaches to reduce the manual efforts required in the secure code review process. Despite the progress, current automated approaches on secure code review, including static analysis, deep learning models, and prompting approaches, still face the challenges of limited precision and coverage, and a lack of comprehensive evaluation.

    To mitigate these challenges, we propose \tool, which is an automated secure code review approach based on large language models (LLMs). \tool leverages a novel mixture-of-prompts architecture that incorporates many prompt experts to improve the coverage of security issues. Each prompt expert is a dynamic prompt pipeline to check the existence of a specific security issue. \tool also implements an effective routing algorithm to activate only necessary prompt experts based on the code features in the input program, reducing the false positives induced by LLM hallucination. Experiment results in our internal dataset demonstrate the effectiveness of \tool in security issue identification and localization with an F1 of 63.98\%. The review comments generated by \tool also achieve a high acceptance rate up to 84\% when it is deployed in production environments. 
\end{abstract}

%% file: sections/intro.tex
\section{Introduction}\label{sec:intro}

Code review has been an important step in the software development process to ensure software quality, as it can help identify various software issues, including coding style issues, performance issues, code smells, and security vulnerabilities, at an early stage of software development~\cite{yang24a}.
Among all software issues detected in the code review process, security issues could result in severe financial losses and service disruptions. Therefore, secure code review becomes one of the top goals in the entire code review process of many companies.

Traditional code review
% are conducted by domain experts within the company, who possess in-depth knowledge of software product design and extensive development experience. This process 
is quite time-consuming, as reviewers need to thoroughly understand the functionality and potential impacts of the code. To improve the efficiency of software development, researchers explored many \textbf{static analysis} tools, such as CppCheck~\cite{cppcheck} and Flawfinder~\cite{flawfinder}, which are built to detect flaws and dangerous coding constructs. They typically rely on well-designed static rules to identify potential coding patterns that may lead to issues. Given the large number of categories in security issues,
% they usually require substantial efforts of domain experts to design rules for each security issue, and 
\textbf{static analysis tools are difficult to cover real-world corner cases as well as new issues}~\cite{cppcheck}.
% Additionally, the development process of a static analysis tool is also lengthy when attempting to address new security issues, as domain experts must consider numerous real-world corner cases. 
% This motivates us to consider the learning-based approaches.

Recently, deep learning techniques, such as CodeReviewer~\cite{li22automating} and T5-Review, proposed fine-tuning a model based on the history of code reviews submitted by code reviewers. The fine-tuned models then take the submitted code as input and generate corresponding code review comments.
However, researchers identified that developers commonly overlooked security-related issues when reviewing code in open-sourced projects~\cite{esemYuFLTS23, di_Biase2016Security}.
For example, Biase et al.~\cite{di_Biase2016Security} found that only approximately 1\% of the review comments are related to security.
This indicates that, in real-world software development, it is hard to collect sufficient high-quality code examples for each security issue.
% For example, we have collected 20,044 past code review comments from a production line at our company and found that only 291 of them contain enough information for the models to understand the security issues.
Therefore, \textbf{fine-tuned code review models cannot be quickly adapted to new security issues due to a lack of data for these issues}~\cite{Braz22software,yang2024security}.
% Therefore, \textbf{current fine-tuned code review approaches often overlook security-related issues because of the insufficient training data}, which are the top priority in practical software development~\cite{Braz22software,yang2024security}.
% This motivates us to consider the LLM-based approaches.

In the era of large language models (LLMs), some prompt learning techniques are also studied in code review by carefully designing prompts to ask LLMs to detect issues directly.
Theoretically, the prompt-based approaches do not suffer from the insufficient training data of a specific type of security issue; instead, they rely on the prompts from domain experts and the rich knowledge from the LLM.
However, when deploying the prompt-based secure code review techniques in our company, we observe the following challenges:

\begin{enumerate}%[wide=0pt]
    \item \textbf{Limited Precision due to False Positives}. When identifying the categories of the security issues with/without prior knowledge (i.e., related categories knowledge in the prompt), LLMs can alarm with the security issues that are impossible to exist in the code (i.e., hallucination). This leads to lower precision of LLMs when compared with current static analysis techniques. However, an approach with high false positive rate will significantly increase the burden of developers in secure code review, as they have to frequently check and discuss the validity of generated code reviews.
    \item \textbf{Limited Coverage due to False Negatives.} Different security issues may require different analysis of the program state to distinguish and confirm. It is difficult for current LLM-based approaches to cover most real-world security issues with a few fixed prompts. Besides, the performance of LLMs usually drops when it is used in domain-specific data that is rarely included in the training datasets. The limited coverage poses great threats to the code review process as its goal is to ensure the quality of software by identifying as many software issues as possible. The missing security issues will require more efforts to identify in the following software testing process.
    \item \textbf{Lack of Comprehensive Evaluation.}  Currently, there is no gold standard for evaluating the quality of comments proposed by code reviewers or automated code review approaches. For example, prior studies evaluated the performance of their approaches on the code review comment generation task with BLEU and ROUGE-L \cite{tufano22using, fan_exploring_2024}. While useful, such metrics cannot comprehensively reflect the quality of generated review comments. This hinders the direct usage of them in practice.
    % Section \ref{sec_motivation_example} shows an example where the reported security issue needs a returned value, but the code under review does not have a returned value.
    % This motivates us to 
    % If we could statically capture features of a specific security issue category, given a code under review, we could first infer its potential security issue categories based on its features and then apply the prompts of the potential categories to it, reducing false positives.
    %\item Hard to evaluate: Currently, there is no gold standard for evaluating the quality of comments proposed by code reviewers or automated code review approaches.
%For example, prior studies evaluated the performance of their approaches on the code review comment generation task with BLEU and ROUGE-L \cite{tufano22using, fan_exploring_2024}. Although these evaluation metrics can reflect the similarity between the review comment generated by their approaches and the ground truth, the reported performance can be superficial (the results would be different with different wording but the same meaning) and may miss key information (e.g., code smell types are overlooked while the rest part of the review comments are the same, leaving the generated review comment useless). This motivates us to adopt new metrics to evaluate the performance of the proposed approach.
\end{enumerate}

In this paper, we propose \tool, which is a LLM-based secure code review approach built upon a novel mixture-of-prompts architecture. The mixture-of-prompts architecture incorporates many prompt experts, and each of them is a dynamic prompt pipeline designed by experienced developers for one specific security issue. Prompt experts contain rich knowledge from developers to comprehensively identify potential security issues, increasing the coverage of \tool. Given the input program, we propose a novel routing algorithm to select applicable prompt experts for code review by analyzing the code features in the program. This indicates that \tool only activates highly-related prompt experts and avoids false positives by deactivating most irrelevant prompt experts. \tool is also extendable to new security issues by simply adding new prompt experts. To provide a comprehensive evaluation, \tool not only outputs the review comments, but also identify the security issue categories and pinpoint the related locations in the program.

We evaluate \tool in an internal dataset of the company, which contains programs with different security issues identified by code reviewers in the past. Experiment results show that \tool can achieve an F1 of 63.98\% in real-world security issue identification, and an accuracy of 47.58\% in issue localization. It outperforms current approaches for at least 32.11\% in issue identification and 26.51\% in issue localization. Furthermore, more than half of the review comments generated by \tool are instrumental for developers. \tool also significantly improves the acceptance rate of review comments in production lines by at least 36.84\%.

We summarize our contributions as follows.
\begin{itemize}%[wide=0pt]
    \item To the best of our knowledge, we propose the mixture-of-prompts structure for \tool, which is a brand new prompt approach for software engineering tasks.
    \item We design \tool, a LLM-based secure code review approach to improve both the precision and coverage of previous approaches.
    \item Extensive evaluation demonstrates the effectiveness of \tool in identifying and localizing security issues.
    \item \tool has been practically deployed inside the company and used in real-world software development.
\end{itemize}

%% file: sections/mot.tex
\section{Background}\label{sec:mot}

\subsection{Problem Definition}

\input{tables/cwe}

Code review is a common practice in modern software development processes. It is usually formalized as a generation task $p \rightarrow cmt$, which takes the program $p$ as input and outputs natural language comments $cmt$ to indicate potential issues.
% Note that, in real industrial practice scenarios at our company, we not only focus on developers' current code changes but also review the existing codebase.

In this paper, we define secure code review as a multiple-goal task $p \rightarrow (cat,\ loc,\ cmt)$, which takes a program $p$ as input and outputs the category $cat$ of the security issues and its location $loc$, along with a natural language review comment $cmt$ for further explanation. This task is generally more challenging than regular code review, as it also requires the accurate identification of security issues and locations. With identified security issues and locations, we can evaluate the effectiveness of an approach more objectively.

\subsection{Motivating Example to Reduce False Positives}\label{sec_motivation_example}

\begin{lstlisting}[language=C++,caption=A motivating example simplified from a C program inside the company.,label=lst:mot]
int func(JNIEnv *env, jclass clazz, jstring RootPath, jstring DestPath, jbyteArray data) {
    const char *RootPath = (*env)->GetString(env,RootPath, 0);
    const char *DestPath = (*env)->GetString(env,DestPath, 0);
    size_t buffSize = (*env)->GetArrayLength(env,data);
    uint8_t *buff = (uint8_t *) malloc(buffSize);
    (*env)->GetByteArrayRegion(env,data, 0, buffSize, (jbyte *) buff);
    Error ret = OK;
    ret = SaveBufferToFile(buff, buffSize, RootPath, DestPath);
    (*env)->ReleaseString(env,RootPath, RootPath);
    (*env)->ReleaseString(env,DestPath, DestPath);
    if (ret != OK) {
        LogOutput(LOG_LEVEL_ERR, "...", ret);
        return ret;
    }
    free(buff);
    return 0;
}
\end{lstlisting}

\input{tables/mot}

To better illustrate the motivation of our approach, we show an example in Code~\ref{lst:mot}. This example is simplified from a real C program in our company, and it has a \textit{Null Pointer Dereference} issue (i.e., CWE-476) at line 2 and a \textit{Unchecked Return Value to NULL Pointer Dereference} issue (i.e., CWE-690) at line 5, where the external function argument $env$ and the pointer $buff$ are not checked before dereference. Besides, it also has a \textit{Missing Release of Memory after Effective Lifetime} issue (i.e., CWE-401) at line 11, where the program does not release the memory $buff$ when it handles an error. Developers have confirmed these issues, which may lead to severe program errors at runtime.

To understand the performance of existing approaches on this example, we apply several popular approaches to it. Table~\ref{tab:mot} shows the identified CWEs and their locations. Note that we do not include the natural language reviews in the table to save space.

\noindent\textbf{Static Analysis.} Cppcheck~\cite{cppcheck} and FlawFinder~\cite{flawfinder} are two static analysis tools designed to scan potential program flaws for C/C++ programs. We do not select advanced static analysis tools that require a compilation database since the compilation database is usually not accessible in the code review process. We observe that both Cppcheck and FlawFinder do not identify any security issues in this example, while their documentation states that they support the identification of CWE-476 and CWE-401 issues. This indicates the limited coverage of current static analysis approaches.

\noindent\textbf{Code Review Models.} CodeReviewer~\cite{li22automating} and T5-Review~\cite{tufano22using} do not output CWE categories and locations. However, CodeReviewer just generates a comment ``\textit{Please remove this blank line.}'' and T5-Review generates a comment ``\textit{The logic from here would be easier to understand if we used the same `tombstone' as: hasUnusedBuffer(env,DestPath, DestPath)}''. The both comments do not indicate any security issues and useful suggestions.

\noindent\textbf{LLM-based Approaches.} We further evaluate the performance of LLM-based approaches on this example, and design two prompt settings: 1) instructional prompting, where we only query the LLM to list all possible CWE issues, and 2) CWE informed prompting, where we give the LLM the concerned CWE categories (listed in Table \ref{tab:cwe}) and prompt the LLM to check if any of them exist. We employ the prompt settings on two LLMs, Qwen-2.5 and DeepSeek R1, to observe the differences between regular LLMs (e.g., Qwen-2.5) and deep thinking LLMs (e.g., DeepSeek R1).

From Table~\ref{tab:mot} we can see that Qwen-2.5 and DeepSeek R1 both failed to identify the correct CWE issue based on the instruction prompt. This suggests that \textbf{current LLMs are unable to detect the relevant CWE issues without guidance}. However, when given the knowledge of the concerned CWE categories,
% DeepSeek R1 still cannot identify the \textit{Null Pointer Dereference} issue and outputs a false positive. 
Qwen-2.5 can identify the existence of CWE-476 and CWE-401.
\textbf{This indicates that compared with static analysis tools, LLMs have a better performance on secure code review.}
This motivates us to use LLMs on the secure code review task with guidance.

Despite the correct identification, Qwen-2.5 misses the CWE-690 issues at lines 5 and it also provides an incorrect location for CWE-401 issue. Furthermore, it outputs two additional CWE categories, which do not exist in the example code, resulting in a high false positive rate. For example, Qwen-2.5 identifies a CWE-78 issue, but there is no interaction with the system via system calls in the example code. Therefore, \textbf{LLM-based approaches can produce both false positives and false negatives} if we let them examine the relevant issues one by one.

\noindent\textbf{Our Approach.} \tool implements a novel mixture-of-prompts architecture to address current challenges.
It does not prompt the LLMs to check all concerned security issues; instead, it analyzes the features in the code and identifies the pointer dereferences and memory allocation in the example and only activates the prompt expert for checking both issues. Prompt experts for other security issues are not activated and will not be identified by \tool.
By doing so, \tool can minimize the false positives brought by LLMs. Based on the activated prompt experts, \tool can thoroughly examine the existence of related issues and accurately capture all issues and locations, avoiding false negatives.

%% file: tables/cwe.tex
\begin{table}[t]
    \centering
    \caption{The CWE issues covered by \tool in secure code review.}
    \scalebox{0.7}{
    \begin{tabular}{ccc}
    \toprule
        \textbf{Category} & \textbf{SubCategory} & \textbf{CWE Issues} \\
    \midrule
       \multirow{6}*{\tabincell{c}{\textbf{Memory} \\ \textbf{Security}}}  &  Memory Allocation & CWE-131, CWE-401, CWE-789 \\
       & Memory Access & CWE-129, CWE-785. CWE-806\\
       & Memory Release & CWE-415, CWE-416, CWE-590, CWE-761, CWE-762 \\
       & Pointer Dereference & CWE-476, CWE-690, CWE-823 \\
       & Pointer Casting & CWE-587, CWE-588 \\
       & Others & CWE-134, CWE-562 \\
    \midrule
        \multirow{2}*{\tabincell{c}{\textbf{Number} \\ \textbf{Processing}}} & Integer Calculation & CWE-128, CWE-191, CWE-193, CWE-369, CWE-1335 \\
        & Data Size Calculation & CWE-467, CWE-469 \\
    \midrule
        \multirow{2}*{\tabincell{c}{\textbf{Sensitive Info} \\ \textbf{Exposure}}} & Process Exposure & CWE-214 \\
        & Log Exposure & CWE-532 \\
    
    \midrule
        \multirow{2}*{\textbf{DoS Attack}} & Untrusted Data & CWE-502 \\ 
        & Improper Control & CWE-119 \\
    \midrule
        \multirow{2}*{\textbf{Injection}} & Command Injection & CWE-78 \\
        & SQL Injection & CWE-89  \\
    \bottomrule
    \end{tabular}}
    \label{tab:cwe}
\end{table}

%% file: tables/mot.tex
\begin{table}[t]
    \centering
    \caption{The identified CWE issues by different approaches. Correct predictions are highlighted in \textcolor{codegreen}{green} color.}
    \scalebox{0.7}{
    \begin{tabular}{clc}
    \toprule
       \textbf{Approach}  & \textbf{Identified CWEs} & \textbf{Location} \\
    \midrule
        Cppcheck &  None & - \\
    \midrule
        FlawFinder & None & - \\
    \midrule
        CodeReviewer & None & - \\
    \midrule
        T5-Review & None & - \\
    \midrule
        \multirow{4}*{\tabincell{c}{Qwen-2.5 +\\ Instruction Prompt}} & CWE-762 (Mismatched Memory Management Routines) & 15\\
        & CWE-416 (Use After Free) &  9, 10\\
        & CWE-476 (NULL Pointer Dereference) &  12\\
        & CWE-707 (Improper Enforcement of Message or Data Structure) & 5, 6\\
    \midrule
        \multirow{4}*{\tabincell{c}{Qwen-2.5 + \\ CWE Info Prompt}} & CWE-78 ('OS Command Injection') &  8\\
        & \textcolor{codegreen}{CWE-476 (NULL Pointer Dereference)} &  \textcolor{codegreen}{2}\\
        & CWE-532 (Insertion of Sensitive Information into Log File) &  8\\
        & \textcolor{codegreen}{CWE-401 (Missing Release of Memory after Effective Lifetime)} & 15 \\
    \midrule
        \tabincell{c}{DeepSeek R1 + \\ Instruction Prompt} & CWE-670 (Use of Externally Controlled Input) & 2, 3, 4, 5\\
    \midrule
        \tabincell{c}{DeepSeek R1 +\\ CWE Info Prompt} & None & -\\
    \midrule
        \multirow{3}*{\tool} & \textcolor{codegreen}{CWE-690 (Unchecked Return Value to NULL Pointer Dereference)} & \textcolor{codegreen}{5}\\
        & \textcolor{codegreen}{CWE-476 (NULL Pointer Dereference)} &  \textcolor{codegreen}{2} \\
        & \textcolor{codegreen}{CWE-401 (Missing Release of Memory after Effective Lifetime)} & \textcolor{codegreen}{11} \\
    \bottomrule
    \end{tabular}}
    \label{tab:mot}
\end{table}

%% file: sections/meth.tex
\section{Methodology}\label{sec:meth}

\begin{figure*}
    \centering
    \includegraphics[width=1.0\linewidth]{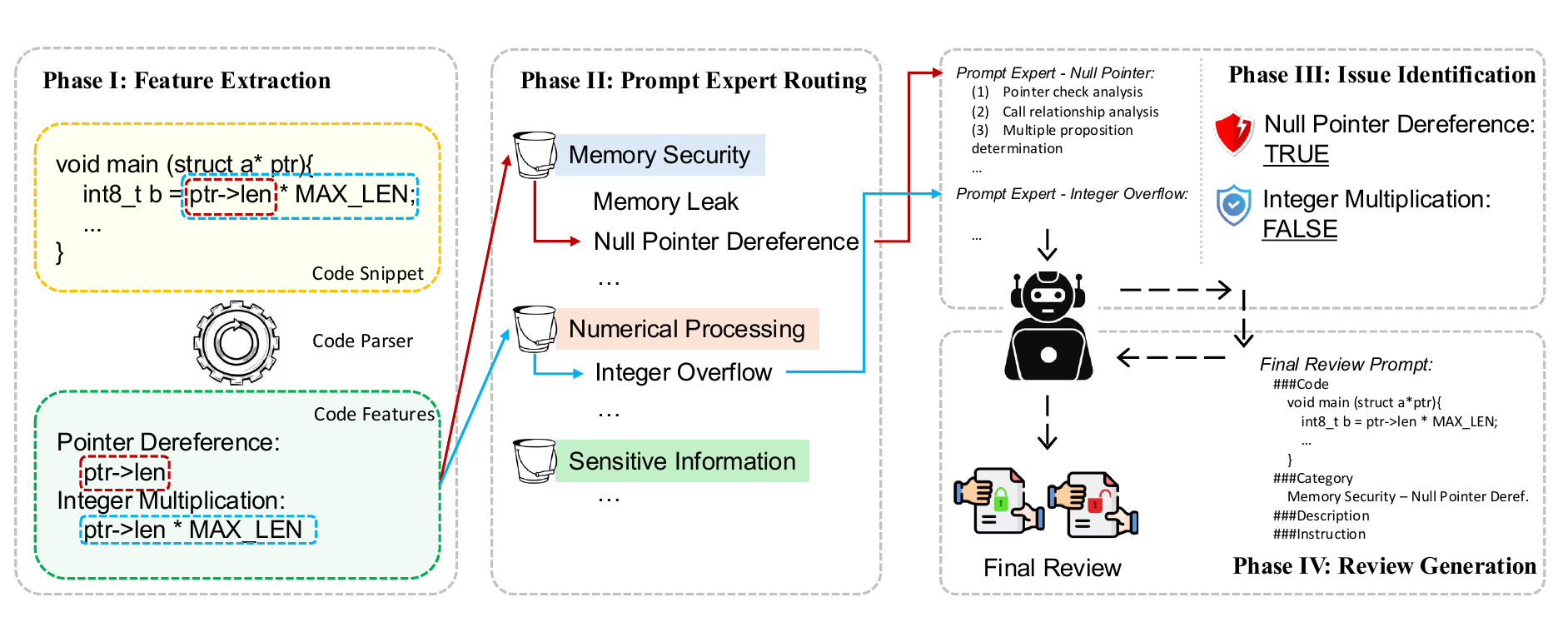}
    \caption{The overview of \tool.}
    \label{fig:overview}
\end{figure*}

\subsection{Overview}

As an LLM-based code review approach, \tool adopts a Mixture-of-Prompts (MoP) approach to particularly detect security issues and provide review comments. 
Fig.~\ref{fig:overview} shows the overall design of \tool that consists of four phases.
Similar to the Mixture-of-Experts (MoE) architecture used in LLMs, \tool hosts a collection of prompt experts, each of which is manually designed by senior developers in our company to identify specific types of security issues (as detailed in Section \ref{sec_method_promptexpert}).
\tool first extracts features of the code under review (Phase I).
These features indicate the existence of potential security issues in the code and then serve as a router in the MoE architecture to select the relevant prompt experts (Phase II). 
The corresponding prompt experts are then activated to confirm whether the corresponding security issue exists or not (Phase III). \tool gathers the identification results from the activated prompt experts and finally generates the final review (Phase IV).

\subsection{Phase I: Feature Extraction}
% Unlike vulnerability detection, which only needs to determine whether a vulnerability exists in the input code, secure code review is more challenging as it requires identifying the exact category of the issue.
% Given this challenge, it is infeasible to activate all prompt experts in a single review, as there are more than 900 CWE categories and over 10,000 special scenarios within our company\jk{but Table \ref{tab:cwe} only have a few...}.
Intuitively, it is unlikely that a code snippet contains all security issues, as the presence of a specific vulnerability is typically associated with certain code features. For example, a code snippet without any memory allocations will never exhibit the memory leak issue. 
Based on this insight, \tool first implements a feature extraction phase to extract the related features in the input code. 

To extract the features, \tool leverages a code parser like tree-sitter~\cite{treesitter} to analyze the input code and transform it into an abstract syntax tree (AST). By traversing the AST, \tool gradually builds a symbol table for all symbols in the code and collects four kinds of features as shown below. 

\textbf{Symbol Table.} 
In most programming languages, an identifier could be regarded as a symbol, and a symbol could be a class, a function, a variable, etc. 
In the AST traversal, \tool collects all symbols and tries to infer the necessary properties of them before building the symbol table. 
To facilitate the secure code review, \tool implements the following three lightweight code analysis techniques to infer the properties of symbols: 
\begin{itemize}%[wide=0pt]
    \item \textbf{Type Inference.} It tries to get the type for each symbol by identifying the definition of each symbol and extracting the declared types. It also infers types for simple expressions and field accesses.
    \item \textbf{Taint Analysis.} It infers whether a symbol is assigned from untrusted sources, such as parameters of functions. 
    \item \textbf{Value Analysis.} It infers the value of a symbol by analyzing initial declarations and direct assignments.
\end{itemize}
It is worth noting that the goal of lightweight analysis is to assist in feature identification, rather than directly detecting security issues. As such, these analyses are not designed to be sound or complete and may fail when essential information is unavailable. \tool does not guess the properties of any symbol to avoid a high false positive rate, and \textbf{leaves the unknown property inference for LLMs in the following phases.} The integration of static analysis and deep learning has been proven effective in program analysis~\cite{peng22static, peng23generative}.

\textbf{Features.} Symbol table records as the state of the input program and cannot be directly used to identify potential security issues.
% To identify potential security issues, we invite domain experts in the company to manually inspect definitions of security issues and determine the related code features.
To identify potential security issues, we invite domain experts from the company to manually inspect the definitions of security issues and instances in the codebase to define code features related to the identification of security issues.
When determining code features, we only include features that can be explicitly verified based on ASTs. For example, for the \textit{Null Pointer Dereference} issue, we only define \textit{pointer type} and \textit{dereference} as two features, rather than the \textit{pointer check}. This is because \textit{pointer type} and \textit{dereference} can be easily identified based on special syntax patterns, whereas \textit{pointer check} is usually implemented using expressions, external APIs, or macros that require complicated analysis to identify.

\tool collects the features of the following four kinds of code based on the symbol table:
\begin{itemize}%[wide=0pt]
    \item \textbf{API}. It identifies all APIs and records the properties of arguments and return values. It also tracks and analyzes the implementation of APIs, provided that their definitions are included in the context.
    \item \textbf{Statement}. It identifies the loop statements and return statements and records the properties of loop variables and returned arguments.
    \item \textbf{Expression}. It identifies the arithmetic and casting expressions and records the properties of operands and results. It also tracks sub-expressions if multiple sub-expressions contribute to an expression.
    \item \textbf{Special Type}. It identifies the special types, including \textit{pointer},\textit{ array}, \textit{container}, \textit{struct}, \textit{macro}, and \textit{class}. It records the properties of all operations involving these special types, such as pointer dereference and array access.
\end{itemize}
For example, in Fig.~\ref{fig:overview}, there are two features extracted from the input code snippet: pointer dereference ``\textit{ptr$\rightarrow$len}'' as it is an operation involving pointer types and integer multiplication ``\textit{ptr$\rightarrow$len * MAX\_LEN}'' as it is an arithmetic expression. The identified code features are then used to activate prompt experts for certain security issues in the next phase.

\subsection{Phase II: Prompt Expert Routing}
Although \tool encompasses a wide range of security issue categories through its prompt experts, it only activates those relevant to the current code context for examination. Based on the features extracted in the first phase, \tool implements a prompt expert routing algorithm to select prompt experts. We present the algorithm in Alg.~\ref{alg:routing}.

\input{algs/matching}

Given the symbol table $ST$ and extracted features $F$, \tool first expands all macros in the code at line 1 to avoid missing critical information in the routing phase. For example, some macros, such as \textit{NODEPTR}, may prevent the identification of pointer types. \tool then classifies all prompt experts into three categories according to their dependencies with the symbol table at line 2. \tool prioritizes and paralizes the prompt experts with no relation with the symbol table ($EP_n$), then handles prompt experts that may read and write the symbol table ($EP_w$), and finally processes prompt experts that only read the symbol table ($EP_r$). This is to prevent the processing of one prompt expert from interfering with another. For each prompt expert, \tool implements both context-free matching and context-sensitive matching to handle both the input program and its associated context.
 
\textbf{Context-free Matching.} For each prompt expert, we design a corresponding matching pattern consisting of two components: 1) a suspicious entity, which is a code element that may lead to security issues if incorrectly operated, and 2) a suspicious operation, which refers to the operations that may be mishandled. \tool activates a prompt expert only when both the suspicious entity and the suspicious operation are detected and matched in the extracted features (line 4-9).
For example, to match the prompt expert for the \textit{Null Pointer Dereference} issue, \tool first locates all pointers in the current program as suspicious entities and all pointer dereference operations as suspicious operations. If any pointer is identified as a suspicious entity involved in a suspicious operation, \tool activates the corresponding prompt expert to conduct further analysis.

\textbf{Context-Sensitive Matching.} \tool follows the same methodology to select prompt experts but implements different techniques to collect suspicious entities and operations. Instead of processing all the context, \tool first identifies the external calls that are related to collected entities and operations (line 10), and then dynamically retrieves the useful code elements from the context based on the issue categories (line 12). For example, in the \textit{Double Free} issue, \tool only identifies the APIs that free the memory in the context and discards all irrelevant contents. This could significantly reduce the code processed by \tool in the routing phase, as the context may sometimes be much larger than the input programs. For the newly collected suspicious entities and operations in the context, \tool matches them back to the symbols in the current program by handling the argument passing in function calls (lines 14-15). The prompt expert will also be activated if a match is found between the new suspicious entities and operations. 

The benefits of the prompt routing phase are twofold: 
1) \textbf{Pruning.} \tool maintains the ability to cover all security issues by keeping all related prompt experts, but it only triggers a few prompt experts in a single review process by filtering out the irrelevant security issues based on the characteristics of the input program.
2) \textbf{False Positive Reduction.} LLMs are not guaranteed to be reliable and could make mistakes due to hallucination~\cite{Huang2025a}. 
The prompt routing phase in \tool leverages lightweight static analysis techniques to reduce false positives by preventing the LLM from being queried about security issues that are not applicable to the given code context.

\subsection{Phase III: Issue Identification}\label{sec_method_promptexpert}

In this phase, \tool sends all activated prompt experts to the LLMs for security issue identification. This phase is parallelized for each prompt expert, as the identification of each security issue is independent. 

\textbf{Prompt experts.} 
Prompt experts are designed to identify the existence of certain security issues, hence they are essential for the performance of \tool. An prompt expert is a prompt pipeline that may contain several sequential prompts ($p_1, p_2, ..., p_n, d$) to infer the security issue, where ($p_1,...,p_n$) is a series of \textit{analysis prompts} aiming to infer the state of the programs, and $d$ is a \textit{determination prompt} that consider the results returned by the analysis prompts and determine the existence of security issues. While different prompt experts have different determination prompts, they may share similar analysis prompts, and the prompt pipeline is dynamically constructed based on the results of previous prompts. Currently, \tool supports the following analysis prompts:
\begin{itemize}%[wide=0pt]
    \item \textbf{Value Inference:} This prompt infers the specific value or value range of a variable.
    \item \textbf{Type Inference:} This prompt infers the type of a variable.
    \item \textbf{Value Check Inference:} This prompt infers all checks implemented for a variable.
    \item \textbf{Taint Variable Inference:} This prompt infers whether a variable is assigned from an untrusted source.
    \item \textbf{Data/Control Flow Path Inference:} This prompt infers the data flow or control flow paths from one point to another.
    \item \textbf{Call Relationship Inference:} This prompt infers the call relationships between the functions to facilitate inter-procedure analysis.
\end{itemize}

The benefit of prompt experts is mainly on the \textbf{false negative reduction}. We use a prompt pipeline instead of a single prompt as a prompt expert to simulate the thinking processes of human reviewers, because security issues are generally more difficult to distinguish and require a comprehensive analysis of program states. Furthermore, we break the design of prompt experts into the combination of analysis prompts and determination prompts, where the analysis prompts act like the ``\textit{shared experts}'' in the MoE structure and can be reused for different security issues, and the domain experts only need to design the determination prompt when adding a new security issue category.

\tool contains 38 prompt experts written by senior developers. It covers five major categories of security issues: \textit{memory security}, \textit{number processing}, \textit{sensitive information exposure}, \textit{DoS attack}, and \textit{injection}. Table~\ref{tab:cwe} presents the security issues that \tool currently supports. For instance, the prompt expert for the \textit{Memory Leak} issue in \tool generally contains three prompts ($p_1, p_2, d$). The first analysis prompt $p_1$ instructs the LLM to analyze the call relationships in the current program and infer the functionality of external APIs. This prompt helps to add missing context information. The second analysis prompt $p_2$ then instructs the LLM to extract the data flow paths from the point of memory allocation to the end of the function. The extracted paths by $p_2$ are then used in the last determination prompt $d$, where the LLM is prompted to determine whether any of them exhibit memory leaks, i.e., at least one path forgets to release the allocated memory.

\textbf{Multiple Proposition Answer for Determination Prompt.} 
Even with the prompt pipeline ($p_1,...,p_n,d$), we still find that LLMs are likely to produce false positives and negatives due to hallucination, especially for security issues with complex identification logic. 
To alleviate this problem, we further design the multiple-proposition answer for complex security issues in the determination prompt $d$. The key insight is to decompose complex identification logic into a series of simpler propositions, requiring the LLM to evaluate each proposition individually rather than directly determining the presence of a security issue. As an example, we design two propositions for the \textit{Integer Overflow} issue: \textit{P1: The result of the operation is used as array index/pointer offset/circulation border/argument of memory allocation/length of memory copy.} and \textit{ 
P2: For any of the operands, there exists a value check.
} 
The LLMs are queried to evaluate the truth values of individual propositions. Then, \tool finally determines the existence of the security issue by computing the values of all propositions, e.g., \textit{(P1 and not P2)} in this example.

In this phase, LLMs are only queried to give simple results of the propositions without explanations. \tool further judges the existence of a security issue based on the results, and discards the security issues that are determined not to be present in the current program. \tool then collects all identified security issues and generates a review to indicate them in the next phase. 

\subsection{Phase IV: Review Generation}

In this phase, \tool aggregates all security issues that LLMs confirm their existence via prompt experts and prepares them for review generation. Based on the security issues identified in the previous phase, \tool prompts the LLM to generate a review comment that includes the categories, locations, and descriptions of each issue. 

When generating the review comment, \tool prompts the LLM to double confirm and rank the identified security issues. Specifically, \tool first provides the LLM with the input program and all identified security issues, and prompts it to explain the reasons. This is similar to a chain-of-thought process that aims to detect potential errors made by the previous phase. The LLM will remove several identified security issues if it finds them unreasonable. \tool then collects the remaining security issues and queries the LLM to generate a complete code review by prioritizing the issues with higher severity. The severity could be either configured by users or determined by the LLM itself.

\tool finally outputs the generated review for the input program, which contains the category $cat$, location $loc$, and descriptions $cmt$ of the identified security issues. Developers and code reviewers can quickly check and verify the correctness of the generated code review by \tool.

%% file: algs/matching.tex
\begin{algorithm}[t]
\caption{Expert Prompt Routing}
\label{alg:routing}
\begin{algorithmic}[1]
\Require
Symbol table, $ST$;
Extracted code features, $F$;
Expert prompts, $EP$;
\Ensure
Selected expert prompts, $P$;
\State $F \leftarrow$ macroExpansion($F$, $ST$)
\State ($EP_n$, $EP_w$, $EP_r$) $\leftarrow$ classifyByDependency($EP$, $ST$)
\For{$ep \in EP_n$ + $EP_w$ + $EP_r$} 
\State \Comment{Context-free matching}
\State $se \leftarrow$ getSuspiciousEntities($F$, $ep$)
\State $so \leftarrow$ getSuspiciousOperations($F$, $ep$)
\If{$se \neq \phi \And so \neq \phi \And$ matches($se$, $so$)}
\State $P \leftarrow P$ + \{$ep$\}
%\State $ST \leftarrow$ updateSymbolTable($ep$, $se$, $so$)
\EndIf
\State $ec \leftarrow$ getExternalCalls($F$, $se$, $so$, $ep$)
\If{$ec \neq \phi$} \Comment{Context-sensitive matching}
\State ($st_c$, $se_c$, $so_c$) $\leftarrow$ dynamicRetrieve($ec$, $ep$)
\State $ST \leftarrow ST $ + $st_c$
\State $se \leftarrow$ matchSymbol($se$, $se_c$)
\State $so \leftarrow$ matchSymbol($so$, $so_c$)
\If{$se \neq \phi \And so \neq \phi \And$ matches($se$, $so$)}
\State $P \leftarrow P$ + \{$ep$\}
%\State $ST \leftarrow$ updateSymbolTable($ep$, $se$, $so$)
\EndIf
\EndIf
\EndFor
\end{algorithmic}
\end{algorithm}

%% file: sections/setup.tex
\section{Experiment Setup}\label{sec:setup}

\subsection{Research Questions}
We focus on the following research questions:
\begin{itemize}%[wide=0pt]
    \item \textbf{RQ1:} How effective is \tool on secure code issue identification compared with existing approaches?
    \item \textbf{RQ2:} How helpful are the review comments generated by \tool in practice?
    \item \textbf{RQ3:} What are the impacts of different components in \tool?
\end{itemize}

\input{tables/benchmark}

\subsection{Datasets}
We evaluate \tool on an internal dataset collected from the code reviews in multiple production lines of the company. The original data contains the locations and review comments. We invite the developers on the corresponding production line to label the related CWE categories. The dataset includes 345 programs with real-world security issues detected by developers and 337 benign programs that developers confirm to be false positives. The dataset consists of 360 C/C++ programs, 181 Java programs, 101 Python programs, and 40 Shell programs. We show the distribution of different issue categories in Table~\ref{tab:benchmark}.

\subsection{Metrics}
As the output of \tool contains three parts: identified issue, location, and review comments, we use different metrics to evaluate its performance. For security issue identification, we follow previous work~\cite{Zou19vuldeepecker} and use multi-class weighted \textbf{Precision}, \textbf{Recall}, and \textbf{F1} to handle the unbalanced distribution of issue categories in our internal dataset. In issue localization, we define \textbf{accuracy} as the ratio of correctly identified and located issues to all security issues in the internal dataset. We define a location as correct if the distance between it and the ground truth is within 1. For review comments, we adopt the methodology from Yu \etal~\cite{yu2024an} and classify all review comments into four categories:
\begin{itemize}%[wide=0pt]
    \item \textbf{Instrumental (I):} The generated review comment explicitly indicates the existence of the security issue identified by the reviewer and provides a fully accurate description.
    \item \textbf{Helpful (H):} The generated review comment raises concerns related to the security issue, but may not be entirely accurate or specific enough.
    \item \textbf{Misleading (M):} The generated review comment does not contain helpful information or has misleading information, such as claiming no security issues are found or reporting a false positive.
    \item \textbf{Uncertain (U):} The generated review comment points out other security issues other than the desired one. Due to a lack of context and knowledge, it is hard to confirm the existence of identified security issues.
\end{itemize}

Based on the above four categories, we evaluate the quality of review comments on the two metrics \textbf{I-Score} = $\frac{I}{I+H+M+U} \times$ 100\%, \textbf{IH-Score} = $\frac{I+H}{I+H+M+U} \times$ 100\%, and \textbf{M-Score} = $\frac{M}{I+H+M+U} \times$ 100\%. We also add a metric \textbf{Acceptance Rate}, the ratio of accepted review comments by developers, to evaluate the practical value of review comments.

\subsection{Baselines}

We compare \tool with the following two widely used static analysis tools:
\begin{itemize}%[wide=0pt]
    \item \textbf{CppCheck}~\cite{cppcheck}: It is a static analysis tool for C/C++ code. It provides unique code analysis to detect bugs and focuses on detecting undefined behavior and dangerous coding constructs.
    \item \textbf{Flawfinder}~\cite{flawfinder}: It scans C/C++ source code and reports potential security flaws.
\end{itemize}

We also evaluate the performance of \tool by comparing it with the following deep learning-based approaches:
\begin{itemize}%[wide=0pt]
    \item \textbf{CodeReviewer}~\cite{li22automating}: It is a model pre-trained with code change and code review data to support code review tasks.
    \item \textbf{T5-Review}~\cite{tufano22using}: It is  a pre-trained Text-To-Text Transfer Transformer (T5) for automated code review.
    \item \textbf{Instruction Prompt}: We use a simple instruction prompt to reflect the basic performance of LLMs by querying them to output all potential CWE issues. 
    \item \textbf{Prompt w/ CWE info}~\cite{yu2024an}: Yu \etal~\cite{yu2024an} evaluate multiple prompting approaches on secure code review and find that prompts with CWE information perform the best. We use it to represent the performance of a general prompt for various security issues.
\end{itemize}

In accordance with the information protection policy in the company, we do not use any closed-source LLMs, such as ChatGPT, as the base models for baselines. Instead, we use the two open-source models Qwen-2.5~\cite{qwen25} and DeepSeek R1~\cite{deepseekai2025deepseekr1} for the prompting approaches to observe the performance of regular and deep thinking LLMs.

\subsection{Implementation}
\tool has two versions for different scenarios of secure code review: 1) a web service to review the code submitted in pull requests, and 2) an IDE plugin to review the code under development. The initial version of \tool is written in Python. \tool uses the tree-sitter library~\cite{treesitter} to parse the code of different programming languages into ASTs in feature extraction and Qwen-2.5 72B~\cite{qwen25} as the base LLM for issue identification and review generation. For DeepSeek R1~\cite{deepseekai2025deepseekr1}, we choose the \textit{DeepSeek-R1-Distill-Qwen-32B} version.

%% file: tables/benchmark.tex
\begin{table}[t]
    \centering
    \caption{The distribution of issue categories in our internal dataset. ``Benign'' indicates that the programs do not contain security issues.}
    \scalebox{0.9}{
    \begin{tabular}{cccccc}
    \toprule
          \multirow{2}*{\tabincell{c}{\textbf{Memory} \\ \textbf{Security}}} & \multirow{2}*{\tabincell{c}{\textbf{Number} \\ \textbf{Processing}}} & \multirow{2}*{\tabincell{c}{\textbf{Sensitive Info} \\ \textbf{Exposure}}} & \multirow{2}*{\tabincell{c}{\textbf{DoS} \\ \textbf{Attack}}} & \multirow{2}*{\tabincell{c}{\textbf{Injection}}} & \multirow{2}*{\textbf{Benign}}\\
         & & & & & \\
    \midrule
         231 & 9 & 51 & 23 & 31 & 337\\
    \bottomrule
    \end{tabular}}
    \label{tab:benchmark}
\end{table}

%% file: sections/eval.tex
\section{Evaluation}\label{sec:eval}

\subsection{Effectiveness of \tool in Security Issue Identification and Localization}

\input{tables/rq1}

To evaluate the effectiveness of \tool in identifying security issues, we compare it with eight baselines on the internal dataset. We do not limit the number of security issues each approach can output to facilitate the evaluation of their precision. As CodeReviewer and T5-Review only output a review comment without issue categories, we use Qwen-2.5 to label the issue categories based on the comments. We randomly sample 10\% of labeled results and find that the labels are 100\% correct. 

\textbf{Issue Identification.} Based on the predictions of each approach, we then calculate the multi-class weighted precision, recall, and F1 for each approach and present the results in Table~\ref{tab:rq1}. From the table, we observe that \tool achieves the highest F1 of 63.98\%, outperforming the best baseline \textit{DeepSeek R1 with instruction prompts} by 32.11\%. This demonstrates the superior performance of \tool in real-world security issue identification. Although CodeReviewer achieves the highest precision of 100\%, its recall is as low as 0.28\%, which indicates that it can hardly capture security issues in practice. \tool still achieves the highest precision among all LLM-based approaches, benefiting from the mixture-of-prompts architecture. Furthermore, \tool can capture the most real-world security issues with a recall of 62.68\%, outperforming the best approach \textit{DeepSeek R1 with instruction prompts} by a large margin.

\textbf{Issue Localization.} Apart from identification, the ability to accurately pinpoint the security issues is also essential to help developers quickly understand the problems. We further calculate the accuracy of locations given by each approach and list it in the fourth column of Table~\ref{tab:rq1}. Note that we exclude CodeReviewer and T5-Review as they do not output location information. The results in the table suggest that \tool can accurately identify and locate the most real-world security issues with an accuracy of 47.58\%. This outperforms the baselines by at least 26.51\%. Moreover, we find that the gap between LLM-based approaches in recall for issue identification and accuracy for localization is generally larger than that of static analysis tools. This indicates that LLM-based approaches cannot pinpoint the locations in some cases, even if they can identify the correct issues.

For both issue identification and location, \tool significantly outperforms current prompting approaches. This demonstrates the effectiveness of the mixture-of-prompts architecture in \tool for secure code review. By routing the input program to only a few prompt experts, \tool can reduce the false positives and identify more security issues with accurate location.

\answer{1}{\tool is effective at identifying and locating real-world security issues with an F1 of 63.98\% and an accuracy of 47.58\%, outperforming the best baseline by 32.11\% and 26.51\%, respectively.}

\subsection{Helpfulness of Reviews Generated by \tool}

\input{tables/rq2}

Based on the issue categories and locations, we could objectively evaluate the performance of existing secure code review approaches. However, the quality of review comments is still essential as they can help developers quickly understand the problems. To evaluate the quality of review comments, we invite software engineers to manually inspect the comments generated by all approaches for our internal dataset. Furthermore, we also deploy \tool in two production lines to observe the changes in the acceptance rate of review comments.

\textbf{Manual Inspection.} We invite three software engineers with at least five years of experience to inspect the comments generated by all approaches manually. We only provide them with the comments and locations so that they can focus solely on the quality of review comments. We ask the engineers to classify all review comments into four categories: \textit{Instrumental}, \textit{Helpful}, \textit{Misleading}, and \textit{Uncertain}, as stated in Sec.~\ref {sec:setup}. Based on the inspection results, we calculate the related metrics and present them in Table~\ref{tab:rq2}. We identify that \tool achieves the highest I-Score of 52.94\%, indicating that more than half of the review comments generated by \tool are indicative and easy to understand.  On the contrary, the best baseline only achieves an I-score of 32.53\%, which is significantly lower than that of \tool. In addition, we find that LLMs with an instruction prompt obtain much lower IH-Score than those with a CWE information prompt. This is opposite to the F1 in Table~\ref{tab:rq1} and suggests that the instruction prompt achieves higher F1 at the cost of producing more misleading information, increasing the burden on developers. 

\begin{figure}
    \centering
    \includegraphics[width=1.0\linewidth]{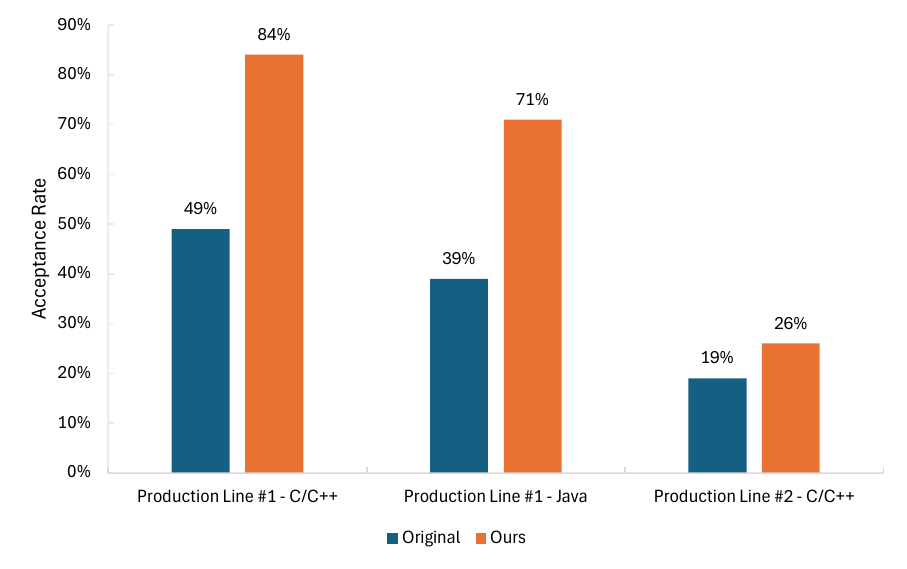}
    \caption{The acceptance rate before and after deploying \tool in two production lines.}
    \label{fig:ac}
\end{figure}

\textbf{Production Line Deployment.} To observe the performance of \tool in real-world software development, we deploy it in two production lines for one week and collect the acceptance rates before and after the deployment. We present the results in Fig.~\ref{fig:ac}. For production line \#1, we can find that the acceptance rate of \tool is 84\% for C/C++ and 71\% for Java, which outperforms the original tool by 71.43\% and 82.05\%, respectively. This suggests that developers accept more than 70\% of reviews generated by \tool in the production lines. As for production line \#2, \tool achieves an improvement of about 36.84\%. However, the acceptance rate in this production line is still low. A possible reason is that the software developed by it is generally more complicated and specific to a single domain, which limits the performance of LLMs.

While better ability in issue identification does not necessarily indicate higher quality of generated review comments, \tool still keeps the good quality of review comments. With the analysis information provided by prompt experts, \tool can generate informative review comments for specific security issues.

\answer{2}{Review comments generated by \tool are helpful with an I-score of 53.94\% and significantly higher acceptance rates in production lines.}

\subsection{Ablation Study}

\input{tables/rq3}

To verify the usefulness of different components in prompt experts, we conduct an ablation study by removing them and observing the performance of \tool. Table~\ref{tab:rq3} presents the results of issue identification and localization for an objective evaluation. For the ablation, we do not remove the routing mechanism in \tool since LLMs with CWE information prompts represent such performance in Table~\ref{tab:rq1} and we have already demonstrated the effectiveness of \tool in RQ1.

In Table~\ref{tab:rq3}, we observe that the highest decrease of F1 from 63.98\% to 47.17\% occurs when we replace all prompt experts with single questions of whether a security issue exists or not. Without the prompt experts, \tool has a similar performance with the instruction prompts. The significant decrease demonstrates that prompt experts are the key component for the mixture-of-prompts architecture. When removing the analysis prompts and multiple proposition answers, the F1 of \tool drops from 63.98\% to 62.23\% and 60.24\%, respectively. This verifies the usefulness of analysis prompts and multiple proposition answers in identifying more security issues. The performance drop in both ablations is not as large as the removal of prompt experts because they are designed only for some complex security issue categories. As for the issue localization, the removal of analysis prompts does not cause a significant decrease since analysis prompts mainly help determine whether a security issue exists. Multiple proposition answers contribute more to issue localization since some propositions indicate location guidance.

\answer{3}{Prompt experts in \tool play the most important roles, and removal of them leads to a significant F1 decrease from 63.98\% to 47.17\% and an accuracy decrease from 47.58\% to 37.89\%. Other components in \tool also contribute to its final performance.}

%% file: tables/rq1.tex
\begin{table}[t]
    \centering
    \caption{Performance of current approaches in security issue identification and localization, in terms of precision, recall, micro F1, and accuracy. `w/ Instr` indicates the instruction prompt and `w/ CWE` indicates the prompt with CWE information. `DS R1` indicates DeepSeek R1.}
    \begin{tabular}{cccc|c}
    \toprule
        \multirow{2}*{\textbf{Approach}} & \multicolumn{3}{c|}{\textbf{Issue Identification}} & \textbf{Localization}\\
    \cmidrule{2-5}
        & Precision & Recall & F1 & Accuracy \\
    \midrule
        CppCheck & 94.36 & 7.69 & 13.17 & 7.41\\
        FlawFinder & 90.16 & 1.14 & 1.20 & 1.14\\
        CodeReviewer & \textbf{100} & 0.28 & 0.57 & - \\
        T5-Review & 94.25 & 2.85 & 5.37 & - \\
        Qwen 2.5 w/ Instr & 73.58 & 41.31 & 45.59 & 30.20\\
        DS R1 w/ Instr & 67.19 & 47.29 & 48.43 & 35.33\\
        Qwen 2.5 w/ CWE & 62.48 & 30.77 & 39.04 & 29.06\\
        DS R1 w/ CWE & 54.84 & 43.87 & 45.25 & 37.61\\
    \midrule
        \tool & 75.48 & \textbf{62.68} & \textbf{63.98} & \textbf{47.58}\\
    \bottomrule
    \end{tabular}
    \label{tab:rq1}
\end{table}

%% file: tables/rq2.tex
\begin{table}[t]
    \centering
    \caption{Helpfulness of reviews generated by current approaches. ``$\uparrow$'' indicates a higher value is better and ``$\downarrow$'' indicates a lower value is better.}
    \begin{tabular}{cccc}
    \toprule
        \textbf{Approach} & \textbf{I-Score} $\uparrow$ & \textbf{IH-Score} $\uparrow$ & \textbf{M-Score} $\downarrow$\\
    \midrule
        CppCheck & 1.80 & 5.58 & 94.42 \\
        FlawFinder & 1.33  & 18.00 & 82.00 \\
        CodeReviewer & 0.31 & 0.31 & 0.70 \\
        T5-Review & 2.96 & 5.62 & 93.50 \\
        Qwen 2.5 w/ Instr & 21.69 & 23.83 & 76.04  \\
        DS R1 w/ Instr & 12.75  & 18.44 & 72.73 \\
        Qwen 2.5 w/ CWE & 32.53 & 40.06 & 59.94 \\
        DS R1 w/ CWE& 21.72 & 33.19 & 65.24\\
    \midrule
        \tool & \textbf{53.94} & \textbf{59.70} & \textbf{40.30} \\
    \bottomrule
    \end{tabular}
    \label{tab:rq2}
\end{table}

%% file: tables/rq3.tex
\begin{table}[t]
    \centering
    \caption{Ablation results of \tool in security issue identification and localization, in terms of precision, recall, micro F1, and accuracy. }
    \begin{tabular}{cccc|c}
    \toprule
        \multirow{2}*{\textbf{Approach}} & \multicolumn{3}{c|}{\textbf{Issue Identification}} & \textbf{Localization}\\
    \cmidrule{2-5}
        & Precision & Recall & F1 & Accuracy \\
    \midrule
        w/o Analysis Prompts &  75.60 & 60.11 & 62.23 & 47.01\\
        w/o Multiple Prop. & 78.22 & 58.69 & 60.24 & 43.87\\
        w/o Prompt Experts &  75.26 & 43.30 & 47.17 & 37.89\\
    \midrule
        \tool & \textbf{75.48} & \textbf{62.68} & \textbf{63.98} & \textbf{47.58}\\
    \bottomrule
    \end{tabular}
    \label{tab:rq3}
\end{table}

%% file: sections/discussion.tex
\section{Threats to Validity}\label{sec:discussion}

\subsection{Internal Validity}
Our study may face the following threats to internal validity.

\textbf{Subjective Evaluation of Reviews.} We evaluate the quality of review comments generated by \tool based on human judgments. This may introduce subjective factors that threaten the validity of evaluation results. We mitigate this threat by using two evaluation methods. We evaluate \tool on the internal dataset by asking several developers to classify the comments into four categories mentioned in Sec.~\ref{sec:setup}-C. Besides, we deploy \tool in two production lines and measure the acceptance rates of code reviews generated by \tool. We believe that the incorporation of different evaluation metrics and developers could significantly reduce the subjective factors and lead to solid evaluation results.

\subsection{External Validity}
Our study may face the following threats to the external validity.

\textbf{Generalization to Other LLMs.} While \tool is a prompt framework and could be implemented upon any LLMs, we only implement it on Qwen-2.5, due to the information protection policy in the company and the limited computation budgets. This may threaten the performance of \tool if it is adopted in other LLMs. However, we believe the adaptation will not significantly hurt the performance of \tool as \tool does not require fine-tuning or use any specific features of the Qwen-2.5 model.

\textbf{Generalization to Other Companies.} \tool is initially built for checking the security issues of our company. We admit that different companies may target different security issues. The performance of \tool may be influenced if it is adopted to check new security issues. However, we believe that \tool could be easily adapted for new security issues by designing new prompt experts and adding them into the prompt router. Besides, the existing security issues supported by \tool are general issues with CWE categories. We do not include any security issues that are specific to our company in the evaluation. 

%% file: sections/literature.tex
\section{Related Work}\label{sec:literature}

% \subsection{Traditional Code Review}

\subsection{LLM-based Code Review}
% TO BE DONE by 28th 
New trends in code review have increasingly focused on leveraging LLMs, due to their flexibility and context-aware reasoning capabilities over both natural and programming languages. 
Tufano et al.~\cite{tufano2022using} started with a fine-tuned T5 model to generate code review comments, showing that pre-trained Transformers outperform prior neural and statistical baselines.
Empirical studies~\cite{cihan2024automated,watanabe2024use} evaluate LLMs such as Codex and ChatGPT in real-world review scenarios. 
They found that while LLMs can replicate many aspects of human review behavior, they may overgeneralize or hallucinate suggestions without project-specific grounding.
Then, researchers also applied agent-based architectures that modularize code review workflows.
For example, CodeAgent~\cite{tang2024codeagent} orchestrates specialized sub-agents (e.g., QA-Checker) to emulate collaborative review dynamics, outperforming baseline generative models such as ChatGPT and Codex.
Unfortunately, we could not include this approach as a baseline due to the political matter.
Further studies~\cite{haider2024prompting,adhalsteinsson2025rethinking,sun2025bitsai} demonstrated that providing additional code structure or execution context improves review accuracy and relevance.
Although additional information is already known to be helpful for code review automation, we are the first to propose the mixture-of-prompts approach that routes the prompt experts based on code features.

\subsection{LLM-adapted Vulnerability Detection}
% TO BE DONE by 27th 
Recent studies proposed LLM adaptation techniques for vulnerability detection. 
There are three major ways to adapt LLMs to vulnerability detection: (1) fine-tuning, (2) prompt engineering, and (3) retrieval augmented generation (RAG). 
During fine-tuning LLMs, researchers~\cite{liu2024pre,peng2023ptlvd,wang2024combining,zhang2023vulnerability,tran2025detectvul,weng2024matsvd} have leveraged various program analysis techniques to extract structural features/relations within code, which can help enhance code understanding.
To address the Transformer~\cite{vaswani2017attention}'s architectural limitations, i.e., sequential token relation, researchers~\cite{tang2023csgvd,jiang2024dfept,yang2024security} tried to apply deep learning modules such as GNN~\cite{scarselli2008graph}. 
There is another study~\cite{ziems2021security} that considers restrictions on the length of input code snippets and applies Bi-LSTM~\cite{schuster1997bidirectional} to mitigate the limitation.
While fine-tuning techniques have shown measurable improvements in detection performance, the gains are often modest and come with substantial computational cost and limited generalizability across diverse codebases.

To further improve the performance, researchers focused on boosting the LLM's capabilities by engineering the prompts.
The possible prompt design considerations are on the Task Descriptions~\cite{zhou2024large,fu2023chatgpt,purba2023software,yin2024pros,zhou2024comparison}, Role Description~\cite{zhou2024large,fu2023chatgpt,khare2025understanding,yin2024pros}, Auxiliary Information~\cite{khare2025understanding,zhou2024large,zhang2024prompt}, and Chain-of-thought~\cite{kojima2022large,zhou2024comparison,ni2024learning}.
Moreover, researchers~\cite{ni2024learning,zhou2024comparison} have also studied the capabilities of prompt design using a few examples of input and ground-truth label pairs.
Unlike these studies, our approach borrows the Mixture-of-Experts (MoE) concept~\cite{jordan1994hierarchical}, which routes the input through specialized sub-models based on the characteristics of the code, and proposes the mixture-of-prompts architecture that does not require model training. 
This allows LLMs to adaptively focus on specific potential security issues initially inferred from the input program, which enhances performance to identify security issues accurately.

%% file: sections/conclusion.tex
\section{Conclusion}\label{sec:conclusion}

In this paper, we first define secure code review, a more challenging but practical and helpful code review task desired by the industry. To complete this task, we propose \tool, a LLM-based approach that leverages a novel mixture-of-prompts architecture to improve both the precision and coverage of previous code review approaches. We compare \tool with eight baselines in an internal dataset of the company for an objective evaluation of security issue identification and localization. We also evaluate the quality of review comments generated by \tool via manual inspection of senior developers and deployment in two production lines. Results demonstrate the effectiveness of \tool in security issue identification and localization, and the helpfulness of review comments generated by \tool. \tool is now adopted by many production lines in the company.

%% file: main.bbl
\begin{thebibliography}{10}

\bibitem{adhalsteinsson2025rethinking}
Fannar~Steinn Aalsteinsson, Bj{\"o}rn~Borgar Magn{\'u}sson, Mislav Milicevic, Adam~Nirving Davidsson, and Chih-Hong Cheng.
\newblock Rethinking code review workflows with llm assistance: An empirical study.
\newblock {\em arXiv preprint arXiv:2505.16339}, 2025.

\bibitem{Braz22software}
Larissa Braz and Alberto Bacchelli.
\newblock Software security during modern code review: the developer’s perspective.
\newblock In {\em Proceedings of the 30th ACM Joint European Software Engineering Conference and Symposium on the Foundations of Software Engineering}, ESEC/FSE ’22, page 810–821. ACM, November 2022.

\bibitem{cihan2024automated}
Umut Cihan, Vahid Haratian, Arda {\.I}{\c{c}}{\"o}z, Mert~Kaan G{\"u}l, {\"O}mercan Devran, Emircan~Furkan Bayendur, Baykal~Mehmet U{\c{c}}ar, and Eray T{\"u}z{\"u}n.
\newblock Automated code review in practice.
\newblock {\em arXiv preprint arXiv:2412.18531}, 2024.

\bibitem{deepseekai2025deepseekr1}
DeepSeek-AI et~al.
\newblock Deepseek-r1: Incentivizing reasoning capability in llms via reinforcement learning, 2025.

\bibitem{di_Biase2016Security}
Marco di~Biase, Magiel Bruntink, and Alberto Bacchelli.
\newblock A security perspective on code review: The case of chromium.
\newblock In {\em 2016 IEEE 16th International Working Conference on Source Code Analysis and Manipulation (SCAM)}, pages 21--30, 2016.

\bibitem{fan_exploring_2024}
Lishui Fan, Jiakun Liu, Zhongxin Liu, David Lo, Xin Xia, and Shanping Li.
\newblock Exploring the {Capabilities} of {LLMs} for {Code} {Change} {Related} {Tasks}.
\newblock {\em ACM Transactions on Software Engineering and Methodology}, page 3709358, December 2024.

\bibitem{flawfinder}
Flawfinder.
\newblock Flawfinder, 2024.
\newblock https://dwheeler.com/flawfinder/.

\bibitem{fu2023chatgpt}
Michael Fu, Chakkrit~Kla Tantithamthavorn, Van Nguyen, and Trung Le.
\newblock Chatgpt for vulnerability detection, classification, and repair: How far are we?
\newblock In {\em 2023 30th Asia-Pacific Software Engineering Conference (APSEC)}, pages 632--636. IEEE, 2023.

\bibitem{haider2024prompting}
Md~Asif Haider, Ayesha~Binte Mostofa, Sk~Sabit~Bin Mosaddek, Anindya Iqbal, and Toufique Ahmed.
\newblock Prompting and fine-tuning large language models for automated code review comment generation.
\newblock {\em arXiv preprint arXiv:2411.10129}, 2024.

\bibitem{Huang2025a}
Lei Huang, Weijiang Yu, Weitao Ma, Weihong Zhong, Zhangyin Feng, Haotian Wang, Qianglong Chen, Weihua Peng, Xiaocheng Feng, Bing Qin, and Ting Liu.
\newblock A survey on hallucination in large language models: Principles, taxonomy, challenges, and open questions.
\newblock {\em ACM Transactions on Information Systems}, 43(2):1–55, January 2025.

\bibitem{jiang2024dfept}
Zhonghao Jiang, Weifeng Sun, Xiaoyan Gu, Jiaxin Wu, Tao Wen, Haibo Hu, and Meng Yan.
\newblock Dfept: data flow embedding for enhancing pre-trained model based vulnerability detection.
\newblock In {\em Proceedings of the 15th Asia-Pacific Symposium on Internetware}, pages 95--104, 2024.

\bibitem{jordan1994hierarchical}
Michael~I Jordan and Robert~A Jacobs.
\newblock Hierarchical mixtures of experts and the em algorithm.
\newblock {\em Neural computation}, 6(2):181--214, 1994.

\bibitem{khare2025understanding}
Avishree Khare, Saikat Dutta, Ziyang Li, Alaia Solko-Breslin, Rajeev Alur, and Mayur Naik.
\newblock Understanding the effectiveness of large language models in detecting security vulnerabilities.
\newblock In {\em 2025 IEEE Conference on Software Testing, Verification and Validation (ICST)}, pages 103--114. IEEE, 2025.

\bibitem{kojima2022large}
Takeshi Kojima, Shixiang~Shane Gu, Machel Reid, Yutaka Matsuo, and Yusuke Iwasawa.
\newblock Large language models are zero-shot reasoners.
\newblock {\em Advances in neural information processing systems}, 35:22199--22213, 2022.

\bibitem{li22automating}
Zhiyu Li, Shuai Lu, Daya Guo, Nan Duan, Shailesh Jannu, Grant Jenks, Deep Majumder, Jared Green, Alexey Svyatkovskiy, Shengyu Fu, and Neel Sundaresan.
\newblock Automating code review activities by large-scale pre-training.
\newblock In Abhik Roychoudhury, Cristian Cadar, and Miryung Kim, editors, {\em Proceedings of the 30th {ACM} Joint European Software Engineering Conference and Symposium on the Foundations of Software Engineering, {ESEC/FSE} 2022, Singapore, Singapore, November 14-18, 2022}, pages 1035--1047. {ACM}, 2022.

\bibitem{liu2024pre}
Zhongxin Liu, Zhijie Tang, Junwei Zhang, Xin Xia, and Xiaohu Yang.
\newblock Pre-training by predicting program dependencies for vulnerability analysis tasks.
\newblock In {\em Proceedings of the IEEE/ACM 46th International Conference on Software Engineering}, pages 1--13, 2024.

\bibitem{ni2024learning}
Chao Ni, Liyu Shen, Xiaodan Xu, Xin Yin, and Shaohua Wang.
\newblock Learning-based models for vulnerability detection: An extensive study.
\newblock {\em arXiv preprint arXiv:2408.07526}, 2024.

\bibitem{peng2023ptlvd}
Tao Peng, Shixu Chen, Fei Zhu, Junwei Tang, Junping Liu, and Xinrong Hu.
\newblock Ptlvd: Program slicing and transformer-based line-level vulnerability detection system.
\newblock In {\em 2023 IEEE 23rd International Working Conference on Source Code Analysis and Manipulation (SCAM)}, pages 162--173. IEEE, 2023.

\bibitem{peng22static}
Yun Peng, Cuiyun Gao, Zongjie Li, Bowei Gao, David Lo, Qirun Zhang, and Michael~R. Lyu.
\newblock Static inference meets deep learning: {A} hybrid type inference approach for python.
\newblock In {\em 44th {IEEE/ACM} 44th International Conference on Software Engineering, {ICSE} 2022, Pittsburgh, PA, USA, May 25-27, 2022}, pages 2019--2030. {ACM}, 2022.

\bibitem{peng23generative}
Yun Peng, Chaozheng Wang, Wenxuan Wang, Cuiyun Gao, and Michael~R. Lyu.
\newblock Generative type inference for python.
\newblock In {\em 38th {IEEE/ACM} International Conference on Automated Software Engineering, {ASE} 2023, Luxembourg, September 11-15, 2023}, pages 988--999. {IEEE}, 2023.

\bibitem{purba2023software}
Moumita~Das Purba, Arpita Ghosh, Benjamin~J Radford, and Bill Chu.
\newblock Software vulnerability detection using large language models.
\newblock In {\em 2023 IEEE 34th International Symposium on Software Reliability Engineering Workshops (ISSREW)}, pages 112--119. IEEE, 2023.

\bibitem{qwen25}
Qwen.
\newblock Qwen2.5 technical report, 2025.

\bibitem{scarselli2008graph}
Franco Scarselli, Marco Gori, Ah~Chung Tsoi, Markus Hagenbuchner, and Gabriele Monfardini.
\newblock The graph neural network model.
\newblock {\em IEEE transactions on neural networks}, 20(1):61--80, 2008.

\bibitem{schuster1997bidirectional}
Mike Schuster and Kuldip~K Paliwal.
\newblock Bidirectional recurrent neural networks.
\newblock {\em IEEE transactions on Signal Processing}, 45(11):2673--2681, 1997.

\bibitem{treesitter}
Tree Sitter.
\newblock Tree sitter, 2025.
\newblock https://tree-sitter.github.io/tree-sitter/.

\bibitem{cppcheck}
CppCheck Solutions.
\newblock Cppcheck, 2024.
\newblock https://cppcheck.sourceforge.io/.

\bibitem{sun2025bitsai}
Tao Sun, Jian Xu, Yuanpeng Li, Zhao Yan, Ge~Zhang, Lintao Xie, Lu~Geng, Zheng Wang, Yueyan Chen, Qin Lin, et~al.
\newblock Bitsai-cr: Automated code review via llm in practice.
\newblock {\em arXiv preprint arXiv:2501.15134}, 2025.

\bibitem{tang2023csgvd}
Wei Tang, Mingwei Tang, Minchao Ban, Ziguo Zhao, and Mingjun Feng.
\newblock Csgvd: A deep learning approach combining sequence and graph embedding for source code vulnerability detection.
\newblock {\em Journal of Systems and Software}, 199:111623, 2023.

\bibitem{tang2024codeagent}
Xunzhu Tang, Kisub Kim, Yewei Song, Cedric Lothritz, Bei Li, Saad Ezzini, Haoye Tian, Jacques Klein, and Tegawend{\'e}~F Bissyand{\'e}.
\newblock Codeagent: Autonomous communicative agents for code review.
\newblock {\em arXiv preprint arXiv:2402.02172}, 2024.

\bibitem{tran2025detectvul}
Hoai-Chau Tran, Anh-Duy Tran, and Kim-Hung Le.
\newblock Detectvul: A statement-level code vulnerability detection for python.
\newblock {\em Future Generation Computer Systems}, 163:107504, 2025.

\bibitem{tufano22using}
Rosalia Tufano, Simone Masiero, Antonio Mastropaolo, Luca Pascarella, Denys Poshyvanyk, and Gabriele Bavota.
\newblock Using pre-trained models to boost code review automation.
\newblock In {\em 44th {IEEE/ACM} 44th International Conference on Software Engineering, {ICSE} 2022, Pittsburgh, PA, USA, May 25-27, 2022}, pages 2291--2302. {ACM}, 2022.

\bibitem{tufano2022using}
Rosalia Tufano, Simone Masiero, Antonio Mastropaolo, Luca Pascarella, Denys Poshyvanyk, and Gabriele Bavota.
\newblock Using pre-trained models to boost code review automation.
\newblock In {\em Proceedings of the 44th international conference on software engineering}, pages 2291--2302, 2022.

\bibitem{vaswani2017attention}
Ashish Vaswani, Noam Shazeer, Niki Parmar, Jakob Uszkoreit, Llion Jones, Aidan~N Gomez, {\L}ukasz Kaiser, and Illia Polosukhin.
\newblock Attention is all you need.
\newblock {\em Advances in neural information processing systems}, 30, 2017.

\bibitem{wang2024combining}
Huanting Wang, Zhanyong Tang, Shin~Hwei Tan, Jie Wang, Yuzhe Liu, Hejun Fang, Chunwei Xia, and Zheng Wang.
\newblock Combining structured static code information and dynamic symbolic traces for software vulnerability prediction.
\newblock In {\em Proceedings of the IEEE/ACM 46th International Conference on Software Engineering}, pages 1--13, 2024.

\bibitem{watanabe2024use}
Miku Watanabe, Yutaro Kashiwa, Bin Lin, Toshiki Hirao, Ken'Ichi Yamaguchi, and Hajimu Iida.
\newblock On the use of chatgpt for code review: Do developers like reviews by chatgpt?
\newblock In {\em Proceedings of the 28th International Conference on Evaluation and Assessment in Software Engineering}, pages 375--380, 2024.

\bibitem{weng2024matsvd}
Cheng Weng, Yihao Qin, Bo~Lin, Pei Liu, and Liqian Chen.
\newblock Matsvd: Boosting statement-level vulnerability detection via dependency-based attention.
\newblock In {\em Proceedings of the 15th Asia-Pacific Symposium on Internetware}, pages 115--124, 2024.

\bibitem{yang2024security}
Aidan~ZH Yang, Haoye Tian, He~Ye, Ruben Martins, and Claire~Le Goues.
\newblock Security vulnerability detection with multitask self-instructed fine-tuning of large language models.
\newblock {\em arXiv preprint arXiv:2406.05892}, 2024.

\bibitem{yang24a}
Zezhou Yang, Cuiyun Gao, Zhaoqiang Guo, Zhenhao Li, Kui Liu, Xin Xia, and Yuming Zhou.
\newblock A survey on modern code review: Progresses, challenges and opportunities.
\newblock {\em CoRR}, abs/2405.18216, 2024.

\bibitem{yin2024pros}
Xin Yin.
\newblock Pros and cons! evaluating chatgpt on software vulnerability.
\newblock {\em arXiv preprint arXiv:2404.03994}, 2024.

\bibitem{esemYuFLTS23}
Jiaxin Yu, Liming Fu, Peng Liang, Amjed Tahir, and Mojtaba Shahin.
\newblock Security defect detection via code review: {A} study of the openstack and qt communities.
\newblock In {\em {ACM/IEEE} International Symposium on Empirical Software Engineering and Measurement, {ESEM} 2023, New Orleans, LA, USA, October 26-27, 2023}, pages 1--12. {IEEE}, 2023.

\bibitem{yu2024an}
Jiaxin Yu, Peng Liang, Yujia Fu, Amjed Tahir, Mojtaba Shahin, Chong Wang, and Yangxiao Cai.
\newblock An insight into security code review with llms: Capabilities, obstacles and influential factors, 2024.

\bibitem{zhang2024prompt}
Chenyuan Zhang, Hao Liu, Jiutian Zeng, Kejing Yang, Yuhong Li, and Hui Li.
\newblock Prompt-enhanced software vulnerability detection using chatgpt.
\newblock In {\em Proceedings of the 2024 IEEE/ACM 46th International Conference on Software Engineering: Companion Proceedings}, pages 276--277, 2024.

\bibitem{zhang2023vulnerability}
Junwei Zhang, Zhongxin Liu, Xing Hu, Xin Xia, and Shanping Li.
\newblock Vulnerability detection by learning from syntax-based execution paths of code.
\newblock {\em IEEE Transactions on Software Engineering}, 49(8):4196--4212, 2023.

\bibitem{zhou2024comparison}
Xin Zhou, Duc-Manh Tran, Thanh Le-Cong, Ting Zhang, Ivana~Clairine Irsan, Joshua Sumarlin, Bach Le, and David Lo.
\newblock Comparison of static application security testing tools and large language models for repo-level vulnerability detection.
\newblock {\em arXiv preprint arXiv:2407.16235}, 2024.

\bibitem{zhou2024large}
Xin Zhou, Ting Zhang, and David Lo.
\newblock Large language model for vulnerability detection: Emerging results and future directions.
\newblock In {\em Proceedings of the 2024 ACM/IEEE 44th International Conference on Software Engineering: New Ideas and Emerging Results}, pages 47--51, 2024.

\bibitem{ziems2021security}
Noah Ziems and Shaoen Wu.
\newblock Security vulnerability detection using deep learning natural language processing.
\newblock In {\em IEEE INFOCOM 2021-IEEE Conference on Computer Communications Workshops (INFOCOM WKSHPS)}, pages 1--6. IEEE, 2021.

\bibitem{Zou19vuldeepecker}
Deqing Zou, Sujuan Wang, Shouhuai Xu, Zhen Li, and Hai Jin.
\newblock Vuldeepecker: A deep learning-based system for multiclass vulnerability detection.
\newblock {\em IEEE Transactions on Dependable and Secure Computing}, page 1–1, 2019.

\end{thebibliography}
